# Enhancing camera surveillance using computer vision: a research note

## Haroon Idrees and Mubarak Shah
Center for Research in Computer Vision,
University of Central Florida,
Orlando, Florida, USA, and

## Ray Surette
Department of Criminal Justice,
University of Central Florida,
Orlando, Florida, USA

**Enhancing Camera Surveillance using Computer Vision: A Research Note**

**Introduction**

<u>Police Surveillance of Public Spaces</u>.  Historically called the "stake-out", police surveillance has a long history and evidence gained from surveillance has been an important part of investigations for nearly two centuries (Marx, 1988).  Similarly, the use of visual technology by police began in the nineteenth century with the photographing of inmates and evolved to include crime scene photographs as standard police procedures (Buckland, 2001; Norris & Armstrong, 1999a). While both practices became law enforcement mainstays, police surveillance and visual evidence remained separate realms well into the twentieth century (Norris & Armstrong, 1999a).   Surveillance cameras operated by law enforcement are therefore a relatively recent phenomenon and their marriage has moved surveillance from a human based activity to a heavily technological one.

As evolved, police surveillance has two goals: proactively deterring offenders and aiding in investigations.   Initially, the investigative goal dominated and surveillance was aimed at solving crimes, not preventing them.  But as cameras became less expensive and more pervasive, deterrence and risk reduction became important (Kroener, 2014).  Current camera surveillance projects aim to provide some combination of retrospective crime scene analysis, deterrence of future crimes, and facilitation of real-time intervention and force deployment (Haggerty & Gozso, 2005).

It is unknown how many public space surveillance cameras are operated by law enforcement agencies but a 2014 U.S. estimate was about 30 million (Staples, 2014, p. 71 citing Vlahos, 2009).  Despite their limitations surveillance cameras have emerged as a popular law enforcement choice to address crime and security concerns and much of the gap between what was promised and what was delivered has been linked to their rapid adoption (Surette, 2005).[1] The number of cameras installed quickly outpaced the capacity to monitor them and thus to effectively respond to what was visually captured (Piza, Caplan & Kennedy, 2014a, 2014b; Gill et al, 2005; Keval & Sasse, 2010). It is apparent that a weak link in the information chain from camera to police response is the human monitor tasked with watching the screens (Surette, 2005).

Humans as Camera Monitors: A Poor Match. Humans are not particularly good as camera monitors (Hier, Greenberg, Walby, & Lett, 2007; Näsholm, Rohlfing, & Sauer, 2014; Sutton & Wilson, 2004). Surveillance camera monitors are most frequently tasked with general camera monitoring. They sit at a desk before a bank of monitor screens and conduct on-going non-specific assessment of live video feeds. The review of a surveillance camera's archived video is also sometimes required to determine if an event of interest was recorded. In this second task, monitors are asked to search for a specific event, person, or object. Again, the human monitor is often asked to watch hours of video. The monitoring difficulty is further compounded in that many criminal activities have subtle precursors that are easily overlooked when humans are tasked with monitoring multiple cameras (La Vigne et al, 2011; Piza Caplan & Kennedy, 2014a; Piza & Sytsma, 2016). Humans quickly become image swamped, missing more than they observe even when vigilant (Boksem, Meijman & Lorist, 2005; Faber, Maurits & Lorist, 2012; Gill et al 2005; Sasse, 2010; Surette, 2005).

The deficiency of human monitors occurs because perception failure occurs when there is little visual change present in long video stretches. The monitor's attention shifts from visual review to other non-visual tasks such as conversing or daydreaming resulting in "inattentional blindness" (Johnston, Hawley, Plewe, Elliott, & DeWitt, 1990; Sasse, 2010). In these instances, monitors have their eyes open and are looking at a video stream but their minds are cognitively elsewhere, the visual images failing to reach psychological "attention capture" levels necessary for effective monitoring.[2]  Significant amounts of time and video can pass, the images passing in plain view but unseen (Driver, 1998).

Relevant for specific event searches,  perceptual blindness more often occurs when a monitor's cognitive attention is focused on finding one type of activity to the exclusion of other significant events (Bredemeier & Simons, 2012; Fougnie & Marois, 2007; Most, Scholl, Clifford & Simons, 2005).  In this situation, unexpected and even bizarre events are more likely to fail to capture the attention of monitors.  Important for noting anomalies, such as unexpected crimes, when a monitor is looking intently for a particular element in a video stream, failure to see other things of interest increases (see for example Piza, Caplan, & Kennedy, 2014a, p. 10).   The more different the unexpected event is from what is being looked for, the more likely it is to be missed (Memmert, 2006).[3]  Hence, different but serious crimes than one being searched for are less likely to be noticed, the opposite of the case in general monitoring tasks where the lack of visual change contributes to monitor error.  In addition to these cognitive barriers, a number of surveillance barriers that reduce potential deterrent effects from surveillance cameras have been described.  In addition to defensive actions taken by offenders (Piza & Sytsma, 2016) and camera related contextual factors (Lim & Wilcox, 2016), two additional noted barriers are high camera to operator ratios (Piza, Caplan, Kennedy, & Gilchrist, 2015; Piza, Caplan & Kennedy, 2014a, 2014b) and poor integration into agency practices (La Vigne, Lowry, Markman, & Dwyer, 2011; Piza, Caplan & Kennedy 2014b; Piza, Caplan, Kennedy, & Gilchrist, 2015).

The cumulative result is that "a high camera-to-operator ratio has the predictable result of crime occurring within sight of a camera going undetected and the detection of criminal events by CCTV operators as rare" (Piza, Caplan, & Kennedy, 2014a, p. 1019-1020 citing Norris & Armstrong, 1999a, 1999b). Faced with significant competition for attention, the camera systems currently are 'hit or miss' tools regarding the detection of on-going incidents and expensive time and human capital consuming drudgery-laden search platforms for finding useful investigative evidence. Additionally criticized for using sworn personnel as monitors and for instances of monitor abuses such as voyeurism and profiling, for police agencies the need for an alternative to human monitors is apparent (Bredemeier & Simons, 2012; Surette, 2005). Computer vision applications can potentially address these issues and increase the deterrent impact of cameras and their organizational benefits. However, the lack of computer vision use in law enforcement is exacerbated by a lack of computer vision software development designed with law enforcement needs in mind and the absence of field trials to justify agency costs for upgrading to computer vision enhanced camera networks.

Computer Vision as Solution. An emerging approach to the shortfalls of human monitored camera systems is computer vision (also known as machine vision). A literature search reveals little current use of computer vision capabilities by law enforcement agencies although calls for its incorporation and discussions of potential applications have been forwarded (see Baldwin & Baird, 2001; Barett, Todd, Miller & Blythe, 2005; Piza, Caplan, & Kennedy, 2014b; Shah, Javed & Shafique, 2007; Thomas & Cook, 2006). The bulk of law enforcement computer applications have concentrated not on computer vision but on data-mining coupled with crime mapping to identify crime 'hot spots' (Lohr, 2012; Wang, Ding, Lo, Stepinski, Salazar & Morabito, 2013; Yu, Ward, Morabito & Ding, 2011). The common current computer vision uses are facial recognition applications and license plate readers. While computer vision as a public safety tool remains under-explored, the recent coupling of surveillance cameras to fast, inexpensive computers have made computer vision solutions feasible. The primary benefit that computer vision offers law enforcement agencies is the substitution of automated analysis of camera video streams for human monitors. With computer vision, the human in a computer vision enhanced security camera network assumes a supervisory assessment and response decision role.

The first step in understanding computer vision involves comprehending digitization of a visual image into a grid of pixels where each pixel is assigned a numerical value representing its color. This initial process generates for an image (or in the case of a video, each frame) a two-dimensional grid of numbers which mathematically renders the original image as digits, hence the term 'digital photo'. A simplified example is provided in Figure 1. These assigned numbers are the foundation for all subsequent manipulation, analysis, interpretation, labeling and other higher-level computer vision capabilities. When a digital photo is opened for viewing, the process is reversed by a photo processing program which uses each pixel's value to instruct the viewing device (a computer, smart phone, or digital camera) on how to color a corresponding screen pixel – converting numbers back to colored pixels and reconstructing the picture in a form that humans can see. This "picture to number to picture" process makes computer vision possible.

Figure 1 about here

The key to computer vision is the analysis made possible by the pixel values when the state of the image is not visual but numerical. The art of computer vision moves quickly from input that looks loosely analogous to an 'image' (for example, the numbers assigned to each pixel in Figure 1) to working with data and outputs that do not appear to correspond to the original image in any straightforward fashion as the photo and histogram in Figure 2 demonstrate.

Figure 2 about here

In essence, computer vision involves determining what a quantitative analysis of pixel values can tell about an image. From its numerical foundation the core tasks of computer vision proceed and in turn allow the development of common computer vision applications such as locating people and places in images, face recognition, and image stabilization.

An example of a core computer vision task with direct criminal justice applications is the tracking of objects across a set of video frames (Yilmaz, Javed, & Shah, 2006). The mathematical representations of the tracked objects are derived from determining key points (unique sets of pixels in an image) so that a "probability distribution function" (PDF) can be generated. A PDF is analogous to a visual fingerprint without being as individually unique. Thus, an object will usually have a similar PDF that can be tracked across video frames. The objects tracked can be automatically set by a computer vision object detector or can be assigned manually by a human placing an outline around a region of interest within a video frame. Recent tracking methods can lose and reacquire objects as they move into and out of camera fields of view (Comaniciu, Ramesh, & Meer, 2000) and work well when the tracked object is substantially different from its background. Multiple objects that are similar in appearance or that cross in front or behind one another are more difficult to track.[4]

Another useful computer vision task is the assignment of a name to objects and actions. Not only is it useful to be able to name objects in a picture but it is an important goal to determine whether a human is present and to determine who they are and what they are doing, a crucial public safety surveillance task. For a computer vision program to be able to recognize and name objects, a set of images are initially used to train 'classifiers', computer vision sub-routines that assign labels to images.  Once developed, classifiers can be used to answer queries about unlabeled images such as: Is there a handgun in this video?  The classifier training process for a previously unknown object proceeds after manual annotation during which training data is created by humans who assign labels to a set of representative images of the object (positive visual examples as shown in Figure 3) as well as those which do not contain the object (negative examples). The training images allow the computer vision program to mathematically differentiate among objects. For example, providing examples of "weapons" and "not weapons" trains a classifier that can then better calculate the probability of a weapon being in an image. Such an approach is termed 'supervised' as it assumes availability of annotated training data in contrast to 'semi-supervised' and 'unsupervised' approaches that require partially-annotated or no training data, respectively. In general, the performance of a particular computer vision task is proportional to amount of human-labeled data available.

Figure 3 about here

The now common task of matching a particular individual's face with a face in an image database is also useful (Turk & Kentland, 1991).  To accurately match a face, the computer vision program must consider "between class variation" (different people who share features such as blue eyes) and "within class variation" (the same person who looks differently due to differences in image aspects (frontal face compared with profile for example).  As with object and action labelling, face recognition is set as a probabilistic outcome with some threshold level of image similarity needed to be reached before a match is declared.  A recent improvement works to maximize the distinction between the faces of different persons and takes into account differences (i.e., normalizes within class variation) observed across multiple images of the same person.[5]  With these and other capabilities under development, potential computer vision solutions for police surveillance camera tasks are now within reach.

Computer Vision Applications in Policing.  General law enforcement surveillance needs that computer vision can address fall under two umbrellas.  The first set revolves around the need for automated real-time, live video stream analysis.  The second involves the need for post-hoc searches of archived video files. Computer vision based automated identification of public safety events of interest addresses the first task and query-based searches of video files addresses the second.

Related to the first task, live real-time video analysis involves the need to rapidly identify and correctly respond to ongoing incidents.  This capability is important because effectiveness of a surveillance system in reducing crime has been linked to real-time intervention. Unless surveillance results in someone showing up to address an observed problem, camera deterrent effects wane (Ariel, 2016; Gill, 2003; Goold, 2004; Piza et al 2014a, 2014b; Welsh & Farrington, 2002). The development of live event analysis is conducted along two computer vision paths: action and event detection of pre-identified events of interest and detection of anomalous new, unanticipated but potentially noteworthy events.  For real-time detection of activities, it is imperative that the system analyzes the surveillance video as it is captured and classifies actions and events as they appear.  Of particular interest to public safety monitors are many activities which occur infrequently and are precursors to criminal activity (for example, 'car hopping' where a person pulls on car door handles as they walk along a street would be a precursor to theft from vehicles).  These activities are more difficult to program because first they are rare and therefore have a limited number of examples available for analysis and second, they can be ambiguous and difficult to define mathematically.  Hence, a murderous assault will likely occur only once in the lifetime of a camera's view-shed but it is crucial that it be noted by a computer vision program and that it be distinguished from one person giving another a vigorous friendly hug.  Humans quickly distinguish the two activities; however, computer vision programs must be quantitatively trained to do so. To be useful, anomaly models also must continuously update and incorporate environmental changes, for instance changes in weather, crowd density, or lightning conditions at different times of the day.

Computer vision can also reduce the immense amount of time currently spent reviewing and searching videos.  Even when it is known that a video contains specific images such as weapons, the minutes or seconds of interest are often buried within hours of output.  A computer vision solution to this issue is query-based searches.  To be useful, query-based searches require search options that permits retrieval of objects with particular properties such as a person with specific height, weight, race, gender, or appearance; or 'objects' such as an item a person was carrying like an umbrella or back-pack.  The ability to submit an object and attribute-based search would significantly reduce the number of irrelevant video clips that an investigator must review.  Independent of specific query-linked searches, it is also useful to have computer vision based video summarization programs for the distillation of videos into shortened but accurate summaries.  An eight-hour video can typically be reduced to an edited 'change only' video lasting minutes (Chen, Wang & Wang, 2009; Evangelopoulos et al, 2009; Gao, Wang, Yong & Gu, 2009).

In another potential use of computer vision, recent criminal justice research has used camera footage to study pre-crime visual cues.  For example, Piza, Caplan, & Kennedy (2014b) and Levine, Taylor and Best (2011) used video footage to examine violence precursors and Moeller (2016) and Piza and Sytsma (2016) searched for correlates of illegal drug sales. Computer vision has the potential to significantly aid these research efforts and increase the use of surveillance videos as a data source. A number of prior research efforts have employed surveillance video as data (see Piza & Sytsma 2016; Piza, Caplan, & Kennedy, 2014a; Sampson & Raudenbush 1999; St. Jean, 2007) but their usefulness has been limited by heavy processing and time demands. Despite having a number of years of video, Piza and Sytsma (2016) had to limit analysis to a single year and 62 incidents due to processing workload; in their study, each minute of video equaled 20 minutes of transcription time. Lastly, as implied by Piza & Sytsma (2016) and Moeller (2016) a set of criminological theories and concepts such as routine activities, environmental crime, crime displacement, and hotspot analysis could benefit from the exploration of computer vision generated data.

　　On-going Research. A National Institute of Justice funded study is underway to address three research questions associated with police use of computer vision (Shah, Idees, & Surette, 2015). In this study computer vision analytics for a large surveillance camera network is being developed and their integration into a Public Safety Visual Analytics Workstation (PSVAW) within a municipal police department will be field tested. The law enforcement targeted computer vision analytics under development include the retrieval of objects, concepts and events (Mazaheri, Kalayeh, Idrees, & Shah, 2015); the localization of actions in long untrimmed videos (Soomro, Idrees, & Shah, 2016); the interactive detection of anomalies without annotated training examples (Zavesky & Chang, 2008); and multiple methods for video summarization (Rodriguez, 2010). The research questions addressed are: what is the accuracy and speed of the analytics; what is the organizational fit of computer vision in a police department; and what is the impact of a computer vision capability on a municipal criminal justice system?

Research Question 1: How well do the computer vision algorithms work in the lab? Computer vision algorithms are being evaluated on standard pre-curated, annotated datasets which are partitioned for training and testing. For many computer vision tasks, prior algorithm accuracy has been high, above 90 percent for easy action recognition datasets. However, for challenging datasets accuracy drops to around 60 percent (Kuehne, Jhuang, Garrote, Poggio, & Serre, 2011), a level that would generate numerous false hits in police applications. The goal is to produce computer vision algorithms that are sensitive enough to not miss significant events but also do not swamp human reviewers with large number of erroneously flagged video clips. A second programming goal is to achieve significant reduction in storage and computational cost over large-scale surveillance video archives. The practical impact for a law enforcement agency would be significant gains in search speed and the ability to search thousands of hours of video data (Ye, Liu, Wang, & Chang, 2013).

A computer vision based method for detection of static concepts and dynamic events is also being developed for object detection such as weapons, police officers, police vehicles; and complex event detection like assaults, thefts, and car crashes. Both use features from deep neural networks for processing images and video frames.[6] In the static concept search, a human can query a single concept such as 'police officer' and the system will return a sorted list of video clips in which the concept 'police officer' appears. The complex event detection categorizes video into broad categorical classes of behaviors beyond a brief appearance of single objects. Thus, more challenging activities can be dealt with and video clips can be robustly classified into events such as 'robberies' and 'assaults'.

Regarding the need for real-time video analysis, computer vision software for live online abnormality detection is additionally being created (Javan, Roshtkhari, & Levine, 2013). The quantitative problem amounts to finding patterns in the digital data that significantly deviate from behaviors previously identified empirically. The detection of abnormal behaviors is a difficult task. First, the quantitative definition of a normal versus abnormal visual pattern is not well defined. Second, normal behavior evolves over time and may change significantly as time passes (for instance, many people walking during daylight versus few people walking during nighttime differ visually but both may be normal activity when it comes to crime detection (Lim and Wilcox, 2016; Moeller, 2016). Third, because abnormal events are rare it is difficult to obtain enough examples to train classifiers.

To cope with these challenges, an online dictionary learning approach to detect abnormalities is being pursued which divides long videos into small non-overlapping meaningful clips. Since these segments are computed based on appearance and motion information, many will contain tracked vehicles and people, which are then compared with existing elements in a dictionary thereby permitting the detection of abnormalities (Tran, Bourdev, Fergus, Torresani, & Paluri, 2015). If the flagged anomaly is deemed a normal event, it is added to the dictionary. This allows the computer program to interactively update and recognize a "new normal" such as when a crowded day time street becomes a sparsely populated nighttime scene. The anomaly detection process flags anomalous events from an unsupervised approach so that labeled training data is not required. The normalcy models will also be unique for each camera, since abnormal behavior may vary by camera across a network.

In addition to detection, a computer vision benefit is the ability to automatically summarize video files and screen out irrelevant information (McCarthy & O'Mahony, 2015). One approach is to use computer vision to identify a small set of suspicious video clips in real time from multiple camera feeds or from a large video archive. A promising approach being pursued is built on semantic indexing (SIN) which uses ideas from deep learning and foreground object detection (Shah, Idees, & Surette, 2015). A temporal action localization (finding an action in long videos) approach automatically decomposes an action into several sub-actions, models each sub-action on appearance and duration into distinct steps, and detects sub-actions in an original untrimmed video. An action event usually consists of a sequence of sub-actions/sub-events in a specific order. For example, a robbery action can be decomposed into person A approaching person B, person A producing a weapon and gesturing at person B, person B holding hands aloft, handing over wallet or phone, and the two separating. The approach for localization automatically discovers the number of sub-actions for each action/event from a set of training videos, registering the point in time the action begins and ends in a video. Once identified, these segments can be flagged for human monitor review and deployment decisions.

A second video summarization approach renders a new video that highlights interesting activities in the original video and skims through redundant information to save viewing time. Along these lines, a hierarchical video summarization method is being created which will first identify small video regions termed supervoxels (regions with similar appearance and coherent motion) based on information such as color and motion. Next, high-level objects of interest such as moving humans or vehicles will be incorporated. These information sources will be combined and the defined video segments matched with previously detected and labeled objects. By detecting interesting regions as well as objects, analysis of human and object interactions are possible (e.g. a theft involving a person in a car or a fight involving multiple people).

Research Question 2: Does computer vision work in the field? If a large automated camera system results in event swamping from the flagging of numerous events for review or has no significant impact on daily agency operations, computer vision's promise will be unmet. To address this issue, a set of events of interest to law enforcement and the design and installation of a computer vision workstation in a municipal police department will be evaluated. Table 1 lists 18 objects, events, and interactions of interest to law enforcement that computer vision algorithms are being developed to detect.

Table 1 about here

Some of the events of interest are rare and have proven difficult to locate sufficient numbers of training examples from police surveillance cameras. In addition, some events occur in conjunction with other crimes or are ambiguous. These events seldom happen without other confounding criminal activity or they are hard to identify by annotators (e.g. injured officer and custody events). Thus, it is difficult to train detectors for such less straightforward events to flag them reliably. Correspondingly, anomaly detection in the real-world assumes greater importance.

In terms of agency impact, the key computer vision field component will be a "Public Safety Visual Analytics Workstation (PSVAW – see Figure 4). The PSVAW will have multiple capabilities ranging from detection and localizing objects in camera feeds, labeling actions and events associated with training data, and allowing query based searches for specific events in videos. It will also be programmed to flag pre-trained criminal and new non-trained abnormal events. Using human monitor feedback, the PSVAW will refine the retrieval parameters and improve its search results over time. After repeating a number of iterations, the PSVAW will create an inductive model to detect new activity of interest in real-time so that an initial anomaly will over time become a computer vision trained, recognized, and labeled event.

Figure 4 about here

Research Question 3.  What are computer vision impacts on a criminal justice system? The   presence of surveillance cameras has been forwarded as both possibly increasing the reporting of events to the police or suppressing citizen guardianship levels (Surette, 2006). Besides issues of loss of privacy, costs, and effectiveness, because computer vision surveillance cameras are expected to catch events that humans would miss, more people may be arrested as the criminal justice net becomes wider and finer (Surette, 2005).

The system-wide impact of a computer vision enhanced camera network will be assessed along two dimensions: its impact on the policing of a community through time series analysis of 'all reported crime' and 'calls for service' data, and the local criminal justice system's use of computer vision generated video for investigations and evidence. Utilizing measures of time spent by human monitors on video requests and processing, flagged events and response times, and use of camera video for investigations and case evidence the system-wide impact of computer vision capabilities will be evaluated.

**Conclusion.**  Computer vision has the potential to address a wide set of problems associated with current public space camera surveillance systems from inappropriate use such as profiling and voyeurism to inherent unintended errors by human monitors.  An automated camera monitor system will not view what it has not been programmed to view and when appropriately programmed will reduce the surveillance gaze from falling on unsuitable subjects. Computer vision systems can also greatly reduce the two sources of error and ineffectiveness in the use of public space surveillance cameras.  A computer algorithm will not become bored or distracted during real-time monitoring so events of interest are less likely to go unseen. Simultaneously, an algorithm will not be so focused on a search for a specific event that other important events go unnoticed.  These benefits are currently being developed and tested in computer vision lab settings.

Computer vision should not be expected to be a panacea for law enforcement's surveillance needs and software gaps remain such as algorithms confidently misidentifying images (Nguyen, Yosinski & Clune, 2014).  Additional shortfalls are due to object size (number of pixels) presenting detection errors in labeling small objects such as guns and tracking can be hampered by the occlusion of people and objects. Hence, following computer vision developments from the lab to the field will be an important step.  The promise of computer vision is that the automation of monitoring can upgrade the current reality of a poorly utilized technology expenditure to a reliable public safety tool.  To already budget conscious, low-on-manpower agencies a field evaluated computer vision capability stands as potentially invaluable.

---

[1] When and how these systems work in specific applications remains under debate (Ariel, 2016; Ariel, Farrar & Sutherland, 2015; Williams, 2007). Indirect evidence suggests that offenders take into account the perceived level of surveillance and the likelihood of intervention when deciding

whether to commit certain types of crimes, especially instrumental street crimes such as car break-ins.  This suggests that easily visible cameras with signage can deter certain offenders (Gill & Loveday, 2003; Allard, Wortley & Stewart, 2008; Short & Ditton, 1996; Welsh & Farrington, 2009).  Spontaneous crimes such as assaults appear to be less affected and overall cameras appear most effective in reducing crime when combined with other interventions (Lim & Wilcox 2016; Piza, Caplan & Kennedy 2014a, 2014b; Welsh & Farrington, 2004).

[2]  Inattentional blindness is defined as the failure to see highly visible objects directly looked at when cognitive attention is elsewhere (Mack, 2003, p. 180; Mack & Rock, 1998; see also Becklen & Cervone, 1983; Neisser, 1979; and Neisser & Becklen 1975).   "Attention capture" refers to the ability of novel stimuli to gain the focus of someone otherwise cognitively engaged (Johnston, Hawley, Plewe, Elliott & DeWitt, 1990; Most, Scholl, Clifford, & Simons, 2005; Wolfe, 1994).  The use of technology has been reported to effect both processes (Hyman, Boss, Matthew, Wise, McKenzie, Kira, & Caggiano, 2009, p. 605) and inattentional blindness has been found to be a common phenomenon associated with watching video streams (Most, Scholl, Clifford, & Simons, 2005).

[3]  A classic example is demonstrated by viewers tasked with counting passes failing to see a gorilla walk through a group of people tossing a ball around.

https://www.youtube.com/watch?v=vJG698U2Mvo

[4] Over the past decade, more sophisticated alternate approaches to tracking have been presented in the computer vision literature, including those that track single and multiple objects or persons across non-overlapping multiple camera field of views (called the 'hand-off' problem) and people in dense crowds (Assari, Idrees, & Shah, 2016; Idrees, Warner, & Shah, 2014; Javed, Rasheed, Alatas, & Shah, 2003; Javed, Rasheed, Shafique, & Shah, 2003).

Figure 1: A Digitized Letter

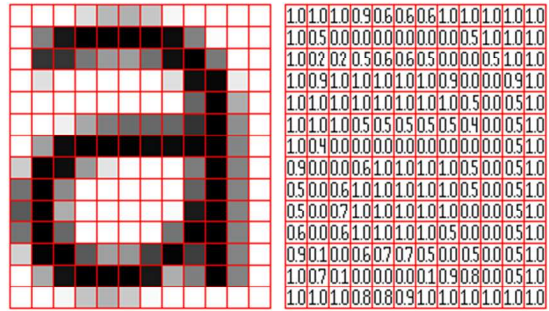

The original letter 'a' is rendered into a 14 by 12 array of 168 pixels with each pixel assigned a value representing a tone from 'white' scored 1.0 to black squares scored 0.0. Gray squares are scored from light to dark (gray squares scored 0.5, light gray ones scored from .01 to .04, dark gray squares scored from 0.6 to .9). The digitized photo of the original 'a' image results where the "a" can be vaguely discerned in the pattern of 0.0 scored pixels. For capturing color, multiple values, typically corresponding to red, green and blue, are stored for each pixel location. Source:
http://pippin.gimp.org/image_processing/images/sample_grid_a_square.png

Figure 2: An Image Histogram

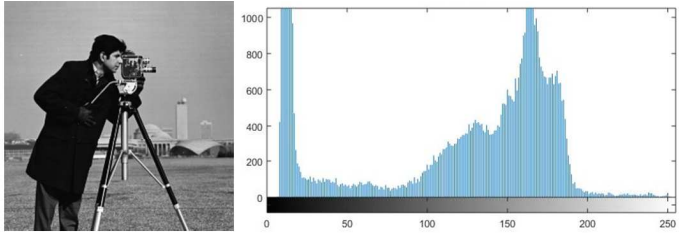

The histogram on the right is constructed from grayscale pixels from the image by quantizing them into 256 bins. The significance of image histograms for computer vision is that histograms are often unique for different objects and correlate with their shape, texture and color and can be used for assigning labels to images.

Figure 3: Positive Training Examples for Assault, Theft, Vandalism, and Robbery.

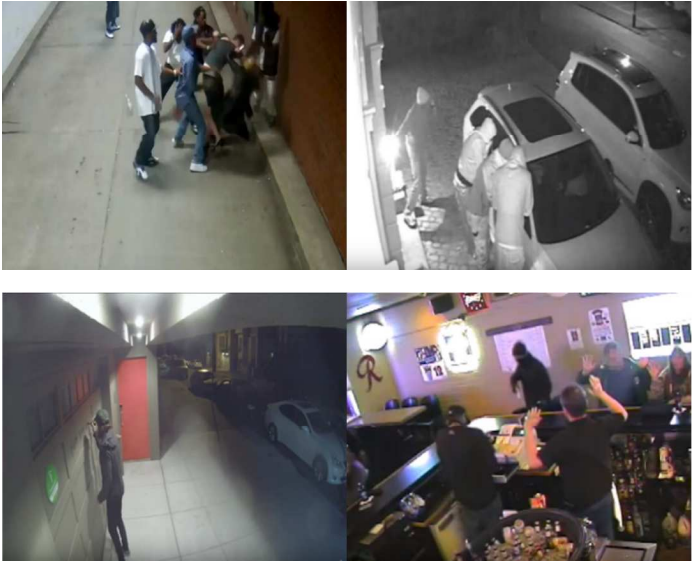

Figure 4:  Public Safety Visual Analytics Workstation (PSVAW)

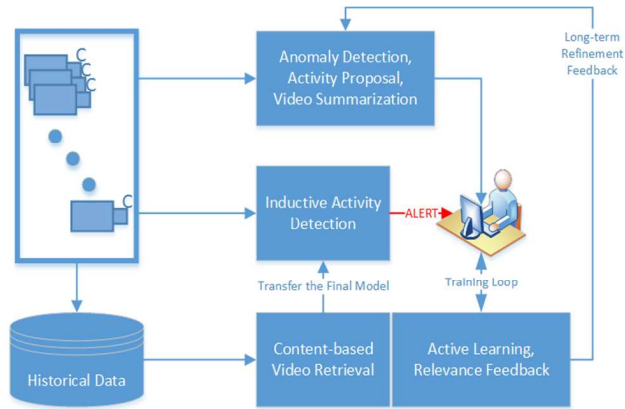

Table 1:   Objects, Crimes, and Events of Interest to Law Enforcement for Computer Vision Algorithm Development and Field Testing.

| Interest | Operational Definition |
|---|---|
| **-- Objects --** | |
| Weapons | handguns, rifles |
| Police Cars | marked law enforcement vehicles |
| Police Officers | uniformed law enforcement officers |
| Emergency Vehicles | fire trucks, ambulances (flashing emergency lights) |
| Left Property | abandoned bags, backpacks, etc., flagged after a set time limit. |
| **-- Crimes --** | |
| Criminal Mischief | graffiti, vandalism (destroying property, tipping over objects, spray painting or other property damage) |
| Theft | removal of property (need to differentiate legal from illegal removal of property such as associated with property damage like breaking a car window) |
| Robbery | need to differentiate legal from illegal exchanges associated with force or weapon |
| Burglary | burglary from auto with window or other property damage; business burglaries constrained to "persons entering a business after closing" possibly flagging constrained from midnight to 6 AM time frame |
| Drug Transactions | need to differentiate legal from illegal exchanges associated with time of day or established illicit drug markets |
| Batteries / Assaults | shootings, stabbings, and fistfights |
| **-- Public Safety Events --** | |
| Car Crashes | damage to moving vehicles |
| Crowd Activity | rioting, crowd density threshold reached, running, fighting, falling, property destruction |
| Individual Activity | running, falling, immobile (exceeds set time limit), blocking traffic, standing in roadway (exceeds set time limit) |
| Injured Officer | officer down (exceeds set time limit) |
| Custody Events | arrests, mental health retentions |
| Citizen / Police | citizen requests for assistance, officer information/identification requests |
| Fire / Explosions | set to minimum size limit. |